\documentclass[aps, prd, twocolumn, nofootinbib]{revtex4}

\usepackage{graphicx}
\usepackage{epic}
\usepackage{eepic}
\usepackage{latexsym}

\newcommand{\eq}[1]{(\ref{#1})}
\newcommand{\be}{\begin{equation}}
\newcommand{\ee}{\end{equation}}
\newcommand{\bea}{\begin{eqnarray}}
\newcommand{\eea}{\end{eqnarray}}
\newcommand{\del}{\partial}
\newcommand{\e}{\epsilon}

\newcommand{\vs}[1]{\vspace{#1 mm}}
\newcommand{\hs}[1]{\hspace{#1 mm}}

\def\b{\beta}
\def\C{\Gamma}
\def\d{\delta}
\def\D{\Delta}
\def\e{\epsilon}

\def\f{\phi}
\def\fr{\frac}
\def\F{\Phi}

\def\k{\kappa}
\def\l{\lambda}
\def\L{\Lambda}
\def\m{\mu}

\def\r{\rho}

\def\del{\partial}

\def\nn{\nonumber}

\begin{document}

\title{Non-Gaussian Path Integration\\ in
  Self-Interacting Scalar Field Theories}  
\author{Ali Kaya}
\email[e-mail: ]{kaya@gursey.gov.tr}
\affiliation{Feza G\"{u}rsey Institute,\\ \c{C}engelk\"{o}y, 81220,
  \.Istanbul, Turkey\vs{2}}
\date{\today}
\begin{abstract}
In self-interacting scalar field theories kinetic expansion is an
alternative way of calculating the generating functional for Green's
functions where the zeroth order non-Gaussian path integral becomes
diagonal in $x$-space and reduces to the product of an ordinary
integral at each point which can be evaluated exactly. We discuss how
to deal with such functional integrals and propose a new perturbative
expansion scheme which combines the elements of the kinetic
expansion with that of usual perturbation theory. It is then
shown that, when the cutoff dependent bare parameters in the
potential are fixed to have a well defined non-Gaussian path integral
without the kinetic term, the theory becomes trivial in the
continuum limit.     
\end{abstract}

\maketitle

\section{Introduction}
In standard model, the scalar Higgs boson is the only fundamental
particle that has not been discovered yet. Our understanding of the scalar
sector is based on rather indirect observations. For instance, one
loop radiative corrections in electroweak theory involve Higgs
particle and the agreement between the theory and the experiment can
be used to estimate the Higgs mass which is expected to be less than
225 Gev \cite{ell}. Also the scalar field is supposed to break
the $SU(2)\times U(1)$ gauge symmetry down to a $U(1)$ subgroup by
acquiring a non-zero vacuum expectation value around 247 Gev
(determined from the experimental value of the W-boson mass and the
Fermi coupling constant). 

Apart from these and some other similar inputs, the scalar sector has
not been  tested directly. There is no experimental data which may be
helpful in  understanding the form and the strength of scalar self
interactions. Specifically, the scalar potential remains to be
determined. Assuming that the interactions are renormalizable and
tractable in perturbation theory, the potential should at most be a
quartic polynomial. Further requiring  spontaneous symmetry breaking,
the potential can uniquely be fixed to have the form  $\lambda
(\phi^2-\mu^2)^2$.    

However, perturbative quantization of scalar fields in standard model
seems unnatural. Namely, there are quantum corrections to Higgs
mass which are quadratic in cutoff $\Lambda$. Therefore, the mass
parameter depends very sensitively on the short distance behavior of
the theory and the value of the bare mass should be very fine tuned to
maintain  Higgs mass within a reasonable range. This is the so called
naturalness or hierarchy problem.  

Moreover, a perturbative renormalization of the
self-interacting $\lambda \phi^4$ theory following the ideas of Wilson
indicates that the theory is free or trivial in the continuum limit
\cite{wil}. Although a non-perturbative investigation of the theory on
the lattice agrees with naive one-loop perturbative result for
renormalization group flow \cite{gr}, it is also possible to argue that in
the large $N$ limit there is a non-trivial fixed point in the four
dimensional scalar field theory \cite{raj}.  

Therefore, it is of great importance to develop new analytical methods 
alternative to usual perturbation theory which may help us to resolve 
these and similar issues in scalar field theories. In this paper,
we consider one such framework, known as kinetic (or strong coupling) 
expansion, where the interaction potential, which can be a
non-polynomial function, is treated  exactly. The main strategy here
is to deal with the kinetic term perturbatively and  carry out
Euclidean path integration in $x$-space (defined with a cutoff), where
the potential is "diagonal" and the functional integral reduces to the
product of an ordinary integral at each point (see
e.g. \cite{st0}-\cite{st7}).  

In this work we first discuss how to calculate a non-Gaussian
functional integral which appears in the zeroth order contribution in
the kinetic expansion. It turns out that the bare parameters in the
potential should be taken cutoff dependent to get a finite result (see
e.g. \cite{stk}). However, the higher order corrections in the
expansion become ill defined due to the presence of uncontrollable
infinities. One can try to perform an extrapolation to cure the
problem as proposed in \cite{st1, st3}, however this procedure likely
fails in general as discussed in \cite{st2,st5}. The main obstacle in
this program is that there is an expansion in derivatives about a
zeroth order configuration which is calculated without paying
attention to locality since the derivative (kinetic) term in the
action is removed \cite{st6}.     

In section \ref{s3}, an alternative expansion scheme is proposed where
one can overcome this obstacle. We find that by introducing a Lagrange
multiplier it is possible to invert the kinetic term  as in a free
theory while the potential can also be integrated out non-perturbatively
(but order by order in cutoff) yielding a complementary series. In
section \ref{s4}, we use this result to show that if the cutoff
dependences of the bare parameters in the potential are chosen
specifically to have a well defined non-Gaussian path integral without
the kinetic term, then the theory becomes trivial in the continuum
limit.        

\section{Perturbative Expansion of the Kinetic Term \label{s2}}

In this section, we mainly review the kinetic expansion of the 
generating functional (see e.g. \cite{st1}).
Consider a self-interacting, real scalar field theory defined in $n$
dimensional space-time. The Euclidean path integral for the generating
functional can be written as   
\be\label{z1}
Z[J]=N \,\int D\phi \,\, e^{\left[\int \b
    \,(\phi\Box\phi)-\,V(\phi)\,+\,J\phi \right]}, 
\ee
where $\int$ means integration over $d^n x$ (the $x$
dependence of functions are suppressed) 
$N$ is the (infinite) normalization, $\b$ is a constant and
$V(\f)$ is the interaction potential including a possible mass
term. We take $n>1$, since we will freely play with the constant $N$
in (\ref{z1}). This is not possible when $n=1$ since in quantum
mechanics the integration measure is uniquely fixed by the
normalization of the state vectors. 

We evaluate \eq{z1} by expanding the exponential of the kinetic term
so that $\b$ is the perturbation parameter. The zeroth order
contribution is 
\be\label{z0}
Z_0[J]=N \,\int D\phi\,\, e^{\left[\int -\,V(\phi)\,+\,J\phi \right]}.
\ee
We first approximate such functional integrals by dividing
$n$-dimensional space into regions of volume $\epsilon$. One can think
that the theory is defined with 
\be
\e=L^n, \hs{6} \Lambda=1/L,
\ee
where $L$ is the spatial lattice size and $\L$ is the cutoff in
momentum space. In this approximation  
\bea
\int f(x)\,d^n x &\to& \sum_i\, f_i \,\,\e,\\ 
\fr{\d}{\d f(x)}&\to& \fr{\del}{\e\,\del f_i}, \label{funcder}
\eea
and path integral measure is given by 
\be\label{dd}
D\phi\to \Pi_i\,\, d\f_i, 
\ee
where the index $i$ runs over the infinitesimal regions. The Dirac
delta function $\d(x-y)$ and the derivative operator can be
represented as $\d_{ij}/\e$ and $(\del \f)_i=(\f_{i+1}-\f_i)/\e$. We
will keep $\e$ infinitesimal but finite and let $\e\to 0$ to reach the
continuum limit. 

Eq. (\ref{z0}) can now be rewritten as
\bea
Z_0[J]&=&N\,\int\,\Pi_i\,d\f_i\,\,
e^{\sum_i\left[-V(\f_i)+J_i\,\f_i\right]\e}\nonumber\\
&=&N\,\Pi_i\,\,z_i(J_i),\label{piz}
\eea
where 
\be
z_i(J_i)=\int_{-\infty}^{\infty} 
d\f_i\,e^{-V(\f_i)\,\e+\,J_i\f_i\,\epsilon}.\label{z=} 
\ee
Thus the problem reduces to the calculation of an ordinary
integral. Although for simple polynomial potentials the exact result
of \eq{z=} can be  
found,\footnote{\label{ft1}
For $V=\l \f^4$ the result of \eq{z=} is 
\bea
&&\frac{\pi}{8\sqrt{2}(\e\lambda)^{3/4}}\times\nn \\
&&\left(\frac{8\sqrt{\e\lambda}}{\Gamma[\frac{3}{4}]}
\,_{0}F_{2}\left[\frac{1}{2},\frac{3}{4};\frac{\e^3J_i^4}{256\lambda}\right]+
\frac{(\e J_i)^2}{\Gamma[\frac{5}{4}]}
\,_{0}F_{2}\left[\frac{5}{4},\frac{3}{2};
\frac{\e^3 J_i^4}{256\lambda}\right]\right), 
\nonumber\eea
where $_{p}F_{q}[{\bf a};{\bf b};x]$ is the generalized hypergeometric
function which is regular at $x=0$ and normalized so that $_{p}F_{q}[{\bf
a};{\bf b};0]=1$. A standard integral table can be found in \cite{boo}} 
for our purposes a series solution in $J_i$ is enough. Of course the
interaction potential should ensure the finiteness of each term and
the convergence of the sum in the series. We take $V$ to be an even
function $V(-\f)=V(\f)$ with $V(\f)\to + \f^p$ for some $p>1$ as
$\f\to\pm\infty$ to satisfy this technical condition. Let us scale 
$\f_i\to\f_i/\sqrt{\e}$ in \eq{z=} to organize the expansion. It is
important to note that any overall multiplicative constant (like
$1/\sqrt{\e}$ which appears after the scaling) can be absorbed in the
definition of the constant $N$ in \eq{piz} and therefore {\it
  ignored}. Introducing  
\be\label{hv}
\widehat{V}(\f_i)=V\left(\f_i/\sqrt{\e}\,\right)\e,
\ee
and
\be\label{asabitleri}
a_m=\frac{1}{m!}\,\,\int_{-\infty}^{\infty}\,e^{-\widehat{V}(u)}\,u^m\,\,du,
\ee
\eq{z=} can be evaluated as
\be\label{expz}
z_i(J_i)=1+\sum_{m=1}^{\infty} \frac{a_{2m}}{a_0}\,\, J_i^{2m}\, \e^m, 
\ee
where $a_0$ is also factored out to ensure the normalization
$z_i(0)=1$. Note that only even powers of $J_i$ survive since we take
$V$ to be an even function.  

Using \eq{expz} in \eq{piz} one obtains
\bea
Z_{0}[J]&=&\Pi_i\left[1+\fr{a_2}{a_0} J_i^2 \e + \fr{a_4}{a_0} J_i^4
  \e^2+...\right],\nn\\  
&=&\exp\left[\sum_i \fr{a_2}{a_0} J_i^2 \e \right] + {\mathcal
  O}(\e),\label{z0e} 
\eea
and in the continuum limit we get
\be\label{z0cont}
Z_0[J]=\exp\,\left[\int\,\fr{a_2}{a_0}\,J(x)^2\,\, d^n x \right].
\ee
Although \eq{z0cont} resembles a functional obtained after a Gaussian
path integration, there is a crucial difference; there are actually
${\mathcal O}(\e)$ corrections to \eq{z0cont} which can be seen in
\eq{z0e}. These can be ignored at this point, however they are not
negligible when $\b$ corrections are calculated (see below).  
 
The constants $a_0$ and $a_2$ might depend on the cutoff so there can
still be infinities hiding in $Z_0[J]$. To get a finite result in the
continuum, $\widehat{V}$ defined in \eq{hv} should not depend on $\e$
as $\e\to0$. This can be ensured by taking the bare parameters in the
potential cutoff dependent.\footnote{Recall that  in usual
  perturbation theory cutoff dependences of the bare parameters in the
  Lagrangian are fixed by requiring finiteness of physical
  quantities. Here we employ the same philosophy.} For instance, if 
\be\label{vv}
V\,=\,m^2\,\f^2\,+\,\l\,\f^p,
\ee
then for $\widehat{V}$ to become cutoff independent $\l$ should be fixed as
\be\label{flow1}
\lambda=\left(\frac{\e}{\e_\m}\right)^{(p-2)/2}
\lambda_\m=\left(\frac{\m}{\L}\right)^{n(p-2)/2}\lambda_\m, 
\ee
where $\e_\m$ corresponds to a fixed scale $\m$, i.e. $\e_\m=\m^{-n}$,
and $\l_\m$ is the ``renormalized'' coupling constant.  Using
\eq{flow1} the cutoff $\L$ and bare coupling $\l$ dependences of $a_0$
and $a_2$ can be traded with $\m$ and $\l_\m$, respectively, which
remain finite in the continuum limit. Note that for the bare mass $m$
there is no need to impose any cutoff dependence. Eq. \eq{flow1} can
be viewed as the analog of the renormalization group flow in this
framework. For $p>2$ the bare coupling vanishes and for $p<2$ it
diverges in the continuum limit $\e\to0$ or $\L\to\infty$. 

To illustrate the above formulas with a specific example, let us take
$V=m^2\f^2+\l\f^4$. Then the constants $a_0$ and $a_2$ can be found 
using \eq{asabitleri} which gives 
\bea
a_0&=&\sqrt{\fr{x}{m^2}}\,e^{x/2}\,K_{1/4}(x/2),  \label{f4a0}\\
a_2&=&\fr{1}{4}\,\left(\fr{\e_\m}{\l_\m}\right)^{3/4} \,\C\left[3/4\right] 
\,_{1}F_{1}\left[3/4,1/2;x\right]\nn\\ 
&-&\fr{m^2}{4}\,\left(\fr{\e_\m}{\l_\m}\right)^{5/4}\C\left[5/4\right] 
\,_{1}F_{1}\left[5/4,3/2;x\right], \label{f4a2}
\eea
where $x=m^4\e_\m/(4\l_\m)$, $K$ is the Bessel and  
$_{p}F_{q}[{\bf a};{\bf b};x]$ is the generalized hypergeometric 
function (see also footnote \ref{ft1}). 
From these, it is clear that  $\l_\m$ dependence of \eq{z0cont} is not
analytic and cannot be obtained in usual perturbation theory.  

The full generating functional \eq{z1} can formally be expressed as
\be\label{fullz}
Z[J]=\exp\left[\b\int G(x,y)\fr{\d^2}{\d J(x)\d J(y)}\, d^n x\, d^n
  y\right]\,Z_0[J]  
\ee
where $G(x,y)=\Box_x\d(x-y)$ and $Z_0[J]$ is given in
\eq{piz}. Eq. \eq{fullz} is ready for a perturbative expansion in $\b$
where the zeroth order contribution is given by \eq{z0cont}. It is
important that one uses \eq{piz} (together with \eq{expz}), but {\it
  not} simply \eq{z0cont}, for $Z_0[J]$ in \eq{fullz} to calculate
higher order corrections. This is because, from \eq{funcder}, the
functional derivatives have a (singular) cutoff  dependence and when
acting on ${\mathcal O}(\e)$ corrections in \eq{z0e} this would yield
non-vanishing contributions in the continuum limit (this is actually
the case as we will see).   

Expanding \eq{fullz} to first order in $\b$ and using \eq{piz} one can find
\bea
Z[J]&=&Z_0[J]+\b \, Z_0[J]\,\,\sum_{ij} G_{ij}\,\fr{\del \ln z_i}{\del
  J_i}\,\fr{\del \ln z_j}{\del J_j}\nn\\ 
&+&\b\,Z_0[J]\,\,\sum_{i}\,G_{ii}\,\fr{\del^2\ln z_i}{\del
  J_i^2}\,+\,{\mathcal O}(\b^2) \label{1c} 
\eea 
where $G_{ij}$ denote the entries of $G(x,y)$ on the lattice. Note
that $\e^2$ factor  coming from the integration measure $d^n x\,d^n y$
in \eq{fullz} is canceled by the cutoff dependence of the functional
derivatives \eq{funcder}. 

Let us now take the continuum limit. Using $z_i$ given in \eq{expz} we
see that the leading order $\e$ terms in \eq{1c} are $\e^2$ and $\e$
for the first and the second lines, respectively, which are exactly
the required powers to convert sums into integrals. Unfortunately,
however, the diagonal entries of $G_{ij}$ diverge for the kinetic term
since in the continuum we have $G(x,y)=\Box_x\d(x-y)$, so there is a
possible divergence problem in the second line in \eq{1c}. 

For the moment, assume that $G(x,y)$ is a smooth and finite
function. Then in $\e\to0$ limit \eq{1c} becomes 
\bea
&&Z[J]=(1+\b \, C) \,\,Z_0[J]\nn\\
&&+ 4\,\b\,\left[\fr{a_2}{a_0}\right]^2\, Z_0[J]\, \int G(x,y)\,
J(x)\, J(y)\, d^n x\, d^n y, \hs{6} \label{g0} 
\eea
where $C=2 (a_2/a_0) \int G(x,x)\, d^n x$ is a constant. The properly
normalized generating functional up to ${\mathcal O}(\b^2)$ is 
\bea\label{g1}
&&\fr{Z[J]}{Z[0]}= Z_0[J]\nn\\
&&+ 4\,\b\,\left[\fr{a_2}{a_0}\right]^2\, Z_0[J]\, \int G(x,y)\,
J(x)\, J(y)\, d^n x\, d^n y, \hs{6}
\eea
so the dependence on a (possibly infinite) contribution $C$ drops out
(not that we set $Z_0[0]=1$). Of course one can now use \eq{z0cont} in
\eq{g1} since there is no functional derivative acting on
$Z_0[J]$. The constant $C$ is related to "vacuum to vacuum" amplitude
and it may also be eliminated by redefining the normalization $N$ in
\eq{piz}. Thus, the first order result \eq{g1} is finite in the
continuum for smooth $G(x,y)$.  

To see what happens for a singular function let us take
$G(x,y)=-m^2\d(x-y)$. Since the mass term can actually be treated
exactly, the perturbative first order result may be derived from
the exact formula by expanding the coefficient of $J(x)^2$ in
\eq{z0cont} around $m=0$ for the potential $V=V_0+ m^2 \f^2$. Using,
on the other hand, the lattice approximation $G_{ij}=-m^2\d_{ij}/\e$
in \eq{1c} one can directly get  
\bea\label{g2}
&&\fr{Z[J]}{Z[0]}= Z_0[J]\nn\\
&&+2\,\b\,m^2\,\left[\fr{a_2^2}{a_0^2}-\fr{6a_4}{a_0}\right]\,
Z_0[J]\, \int \,J(x)^2\, d^n x. \hs{6} 
\eea
As in \eq{g0}, there is a $C Z_0[J]$ term in $Z[J]$ which is
eliminated after dividing by $Z[0]$ where the constant $C$ diverges
like $C\sim \int d^n x$.  

For the kinetic term the
situation is more complicated. Using the approximations for the delta
function and the derivative operator given below \eq{dd} one finds
$G_{ij}=(\d_{i+2,j}\,-\,2\d_{i+1,j}\,+\,\d_{ij})/\e^3$. Therefore
$G_{ii}=1/\e^3$ (no sum in $i$ is implied). However,
it is also possible to define an asymmetric derivative operator such
as $\widehat{\del}\f_i=(\f_{i+1}-\f_{i-2})/(3\e)$ which  gives
$G_{ij}=(\d_{i+2,j-2}-2 \d_{i-1,j}+\d_{i-4,j})/(9\e^3)$ and thus
$G_{ii}=0$. Note that
$(\widehat{\del}\f)_i=(\del\f_i+\del\f_{i-1}+\del\f_{i-2})/3$ so
$\widehat{\del}$ involves an extra averaging over neighboring lattice
points in terms of $\del$. In the continuum limit both $\del$ and
$\widehat{\del}$ asymptote to the same differential  operator. But,
one should actually specify the function space in the continuum more
precisely. Namely, $\widehat{\del}$  can only be used as the
derivative operator if the test functions are not changing appreciably
along three neighboring lattice points. 

To resolve the issue we use the Fourier transform with a momentum
cutoff $\L$ to specify smoothness of the field variables in the
continuum. From    
\be\label{Fr}
G(x,y)=-\int \,\,e^{ip(x-y)}p^2\, d^n p,
\ee
one gets $G(x,x)\sim -\int^\L p^{n+1} dp \sim 
-\L^{n+2}$ and thus we take $G_{ii}=c\,\e^{-(n+2)/n}$ in \eq{1c} where
$c$ is a (finite) numerical constant. In the continuum limit we then
have 
\bea\label{g3}
&&\fr{Z[J]}{Z[0]}=
Z_0[J]\,+4\,\b\,\left[\fr{a_2}{a_0}\right]^2\,Z_0[J]\, \int
\,J(x)\Box_x J(x)\, d^n x \nn\\ 
&&-6\,\b\,c\,\e^{-2/n}\,\left[\fr{2a_4}{a_0}-\fr{a_2^2}{a_0^2}\right]
\,Z_0[J]\,\int  
\,J(x)^2\,d^n x.
\eea 
Unfortunately, the second line of \eq{g3} diverges like
$\e^{-2/n}\sim\L^2$. However, from \eq{g2}, this divergence can be
canceled out by a mass counterterm  $(\d m)^2\sim\b\,\L^2$. So we
conclude that, to first order in $\b$, only a mass renormalization is
required after which all Green's functions become finite.  

As for the higher order $\b$ corrections, it is not difficult to
obtain the analog of \eq{1c} on the lattice. For smooth and finite
$G(x,y)$, ${\mathcal O}(\b^2)$ contribution turn out to be finite in
the continuum. However, analyzing the result for the kinetic term to
this order we find that some Green's functions diverge 
(e.g. quadratically in $\L$) and it seems these cannot simply be 
removed by local counterterms as in \eq{g3}. We also expect more and
more infinities to show up in higher orders coming from the
multiplications of the singular distribution $\Box_x\d(x-y)$. So it
seems that new regularization or renormalization techniques should be
developed for higher orders.\footnote{\label{ftrg} One possibility is
  to set up a renormalization group flow starting from IR. For
  instance, from \eq{Fr} one may decompose    
\bea
G(x,y)=G(x,y,\m)+ \widehat{G}(x,y)\, ,\nn
\eea
where 
\bea
G(x,y,\m)=\,-\, \int_{-\mu}^{\mu}\,e^{ip(x-y)}\,p^2\, d^n p\, ,\nn
\eea
$\m$ is a fixed IR length scale and $\widehat{G}$ contains the
integral in \eq{Fr} starting from $\m$ to $\infty$. Now, in \eq{fullz}
one can first calculate the contribution of $G(x,y,\m)$. Since
$G(x,y,\m)$ is non-singular the exact result of the computation can in
principle be performed to any given order without encountering any
infinities or ambiguities. This gives a new functional $Z_\m[J]$ and
one can now integrate the modes from $\m$ to $2\m$  to obtain
$Z_{2\m}[J]$ etc.} As pointed out in the introduction the main problem
here is that in calculating $Z_0[J]$ locality is neglected since the
kinetic term in the action is removed. On the contrary, the higher order
corrections are sensitive to local variations since more and more
powers of the Laplacian appear in the expansion.  

\section{An Alternative Expansion \label{s3}}

Our main strategy in this section is to use a Lagrange multiplier in the path
integral to separate the kinetic and the potential terms so that \eq{z1}
can be rewritten as 
\be
Z[J]=\int D\phi\,D\F\,D\rho \,\, e^{\left[\int \b
    \,(\phi\Box\phi)-\,V(\F)\,+\,J\F\,+ \,i \rho (\f-\F) \right]}\,
\label{lag} 
\ee
where the normalization $N$ is suppressed. Evidently
integrating over $\r$ gives a delta functional and a further trivial
integration over $\F$ gives \eq{z1}. However one can now carry out the
Gaussian $\f$ integration to get 
\bea
Z[J]=\left. \int D\F D\rho \,\,e^{\left[\int \fr{1}{4\b}
    (\r\Box^{-1}\r)-V(\F)+J\F+ i \rho (K-\F) \right]}\,\right|_{K=0}\nn
\eea
where we have introduced an auxiliary field $K$. Since differentiating
with respect to $K(x)$ gives $i\r(x)$, the last equation can be
expressed as  
\bea
Z[J]&=&\exp \left[-\fr{1}{4\b}\int\fr{\d}{\d K(x)}\Box_x^{-1}\fr{\d}{\d
    K(x)}d^n x\right]\times\nn\\ 
&&\left. \int D\F\,D\r\,e^{\left[\int-V(\F)+J\F+ i \rho
    (K-\F)\right]}\right|_{K=0}. \nn
\eea
At this point, $\r$ and $\F$ integrations become trivial where the
first one gives a delta functional and the second one sets
$\F=K$. After these integrations one finds  
\be\label{dual}
Z[J]\,=\,\left.\,e^{\left[-\fr{1}{4\b}\int\fr{\d}{\d
    K}\,\Box^{-1}\,\fr{\d}{\d K}\right]}\,e^{\left[\int-V(K)\,+\,JK\right]}
\,\,\right|_{K=0}, 
\ee
which can be evaluated order by order in $1/\b$ by expanding the first
exponential. Comparing to \eq{fullz} where $\b$ is the expansion
parameter, we see that \eq{dual} gives a complementary  series,
reminiscent of a strong-weak coupling duality.

It is not difficult to show that for polynomial potentials \eq{dual}
is equivalent to usual perturbation theory. Noting that the
functional derivative of $\exp(\int JK)$ with respect to $J(x)$
($K(x)$) gives $K(x)$ ($J(x)$), \eq{dual} can be rewritten as
\bea
Z[J]&=&\left.\,e^{-\fr{1}{4\b}\int\fr{\d}{\d
    K}\,\Box^{-1}\,\fr{\d}{\d K}}\,e^{\int-V(\fr{\d}{\d J})}\,e^{\int\,JK}
\,\,\right|_{K=0}, \nn\\
&=&\left.\,e^{\int-V(\fr{\d}{\d J})}\,e^{-\fr{1}{4\b}\int\fr{\d}{\d
    K}\,\Box^{-1}\,\fr{\d}{\d K}}\,e^{\int\,JK}
\,\,\right|_{K=0}, \nn\\
&=& \left. e^{\int-V(\fr{\d}{\d J})}\,e^{-\fr{1}{4\b}\int
  J\,\Box^{-1}\,J}\,e^{\int\,JK} 
\,\,\right|_{K=0},\nn\\
&=& e^{\int-V(\fr{\d}{\d J})}\,e^{-\fr{1}{4\b}\int
  J\,\Box^{-1}\,J},\label{usual} 
\eea
where in the second line we have used the fact that $J$ and $K$
derivatives commute and in the last line $\exp(\int JK)$ is dropped
since any $J$ derivative acting on it vanishes after setting $K=0$. This
proves the equivalence since \eq{usual} is nothing but the usual
perturbative expansion formula. From the commutator 
\be
\left[\int \fr{\d}{\d K} \Box^{-1}\fr{\d}{\d K},
  K(x)\right]=2\,\Box_x^{-1}\,\fr{\d}{\d K(x)}, 
\ee
one can also directly verify that \eq{dual} obeys Schwinger-Dyson equation 
\be
2\b\,\Box_x\fr{\d Z[J]}{\d J(x)}-V'\left(\fr{\d}{\d J(x)}\right) Z[J]+
J(x) \,Z[J]=0, 
\ee
where $V'=dV/d\f$. Note that in \eq{dual}, for a fixed order in
$1/\b$ (and for polynomial potentials), only a finite number of terms
survive if one expands $\exp(-\int V(K))$ since one sets $K=0$ at the end.   

Eq. \eq{dual} is preferable over \eq{usual} for organizing the 
perturbative expansion in powers of $1/\b$. Another advantage is that
non-polynomial potentials can also be treated naturally. Assuming that
$V(K)$ is an infinitely differentiable smooth function, \eq{dual}
shows that there is an interaction vertex for the $p$'th derivative of
$V(K)$ if $d^p V/ dK^p\not=0$ at $K=0$. One may reach the same
conclusion in perturbation theory by expanding $V(K)$ around
$K=0$.  

Having rederived the usual perturbation  
theory in a different way, we now
proceed by obtaining another expression for the generating
functional. Let us start with  \eq{lag} in which the external  current
$J$ is coupled to $\f$ rather than $\F$. Integrating over $\F$ one
gets 
\bea
Z[J]=\left. \int D\phi\,D\rho \, Z_0[i\r]\, e^{\left[\int \b
    \,(\phi\Box\phi)+\,J\f\,+ \,i \rho (K-\f) \right]}\right|_{K=0}\, ,\nn
\eea
where the functional $Z_0$ is defined in \eq{z0} and we again
introduced an auxiliary field $K$. Using the same trick we employed
below \eq{lag}, we find 
\be
Z[J]=\left.Z_0\,\left[\fr{\d}{\d K}\right]\,e^{\left[\int \b \,(K \Box
    K)+\,JK\,\right]} \right|_{K=0} 
\ee
Adding and subtracting $\int J\Box^{-1}J/(4\b)$ term in the
exponential a complete square can be obtained. Shifting $K$ as
$K(x)\to K(x)-\Box_x^{-1}J(x)/(2\b)$ one finally reaches 
\be\label{zk}
Z[J]=\left. e^{\left[-\fr{1}{4\b}\int
    J\Box^{-1}J\right]}\,Z_0\left[\fr{\d}{\d K}\right]\, e^{\left[\b
    \int K\,\Box\, K \right]}\,\right|_{2\b \Box K= J}
\ee
or more conveniently 
\be
Z[J]=e^{\left[-\fr{1}{2}\int J\Box^{-1}J\right]}\,Z_0\,\left[\Box
  \fr{\d}{\d J}\right]\, e^{\left[\fr{1}{2}\int J\Box^{-1}J\right]}\, 
, \label{zson}
\ee
where we set $\b=1/2$ since it can no longer be used as an expansion
parameter (note $1/\b$ factors in \eq{zk}). Unlike the expression given
in \eq{fullz}, we now manage to invert the kinetic term as in a 
Gaussian integral. Using the identity
\eq{id1} below it is easy to verify \eq{zson} in the free theory when the
potential contains only a mass term. Note that apart from the first exponential
(and a sign in the third one) $\log Z_0[\Box\f]$ can be viewed as an
effective ``potential'' added to a free massless scalar field. 

In the continuum limit, the functional $Z_0$ is given in
\eq{z0cont}. However, ${\mathcal O}(\e)$ corrections are not
negligible as before. Therefore, {\it a new}  expansion scheme can be
obtained by evaluating $Z_0$ order by order in $\e$ and using the
result in \eq{zk} or in \eq{zson}.  

Let us proceed by a direct computation of the first two terms in this
expansion (a systematic treatment will be presented in the next
section). From \eq{z0e} we obtain  
\bea
&&Z_0[f]=\exp\,\left[\int\,\fr{a_2}{a_0}\,f(x)^2\,\, d^n x \right] \times\nn\\
&& \left\{1+\k\,\e\left(\fr{a_4}{a_0}-\fr{a_2^2}{2a_0^2}\right)\int
f(x)^4 d^n x+{\mathcal O}(\k^2)\right\},\hs{10} \label{zf}
\eea
where $\k$ is a formal expansion parameter which counts the powers of
$\e$ (more comments on this below). To calculate the contribution of
the exponential we note the identity 
\bea
e^{\left[-\int\,\fr{A}{2}\,\fr{\d^2}{\d J^2}\,\right]}e^{\left[\int
    \fr{J^2}{2B}\,   
 \right]}=e^{\left[\int\fr{J^2}{2(A+B)}\right]},\label{id1}
\eea
which can be verified by setting up a simple Gaussian path
integral. In \eq{id1}, an overall irrelevant (infinite) multiplicative
constant is omitted on the right hand side.  
Using  \eq{zf} and \eq{id1} in \eq{zson} the zeroth order term can be
found as 
\be
Z[J]=e^{\left[\fr{1}{2}\,\int \, J \, P_1\, J\,\right]} +{\mathcal O}(\k),
\ee
where 
\be\label{p1}
P_1(x,y)=\left[\fr{a_0}{2a_2}-\Box\right]_x^{-1}\d(x-y).
\ee
The same equations show that to calculate the next order contribution
one confronts the term 
\be\label{36}
\left\{\int \left(\Box_x\fr{\d}{\d J(x)}\right)^4\, d^n x \right\}\, 
\exp\left[\fr{1}{2}\int J\,\Box^{-1}P_2\, J\right],
\ee
where 
\be\label{p2}
P_2\,\,=\,\,\fr{a_0}{2a_2}\,\,P_1.
\ee
The result of \eq{36} contains the $x$-integral of
\bea
&&3\, (\Box P_2)_{xx}^2 \nn\\
&&+ \,6\, (\Box P_2)_{xx}\,
\int\,P_2(x,y_1)\,P_2(x,y_2)\,J(y_1)\,J(y_2)\,dy_1\,dy_2 \nn\\ 
&&+ \int P_2(x,y_1)P_2(x,y_2)P_2(x,y_3)P_2(x,y_4)\times\nn\\
&&\hs{14}J(y_1)\,J(y_2)\,J(y_3)\,J(y_4)\,dy_1\,dy_2\,dy_3\,dy_4.\label{4j}
\eea
The last line can be dropped in the continuum since, from \eq{zf}, the
whole expression is multiplied by $\e$. For the first two lines, one
needs the numerical value of $(\Box P_2)_{xx}$ which can be found from
the Fourier transform 
\be\label{bp}
\Box P_2(x,y)\,=-\,\int\fr{p^2}{1\,+\,2\,p^2\,a_2/a_0}\,e^{ip(x-y)}\,\,d^n p. 
\ee
When $x=y$ the integral diverges like $\L^n$ 
and thus $(\Box P_2)_{xx}=c/\e$ where $c$ is a finite constant. The
first line in \eq{4j} yields an infinity, however this can be pushed
to order $\k^2$ after dividing $Z[J]$ by $Z[0]$. On the other hand
$1/\e$ factor coming from $(\Box P_2)_{xx}$ in the second line is
canceled by the $\e$ term in \eq{zf}, thus there is no problem in
taking $\e\to 0$  limit. Combining all these, the continuum result
becomes 
\bea
&&\fr{Z[J]}{Z[0]}=e^{\left[\fr{1}{2}\,\int \, J \, P_1\,
    J\,\right]}\,\left\{1\,+\, 6\,c\,\k\, 
\left(\fr{a_4}{a_0}-\fr{a_2^2}{2a_0^2}\right)\right. \times\nn\\
&&\left. \int
P_2(x,y_1)P_2(x,y_2)\,J(y_1)J(y_2)\,dx\,dy_1\,dy_2+{\mathcal O}(\k^2) 
\right\}\nn\\
\label{oneloop}
\eea
which does not contain any infinities.
 
Higher order $\k$ corrections can be calculated in a similar
fashion where one now encounters more coincident functional derivatives
acting in \eq{zson} and there are both connected and disconnected
contributions. Of course, more coincident derivatives give a much
more singular behavior. However, these are multiplied by higher
powers of $\e$ and there is a chance that one gets a finite result in
the continuum as for the first order calculation above. This is
indeed the case as will be discussed in the next section. 

The constant $\k$ is just a formal expansion parameter and one should
actually set $\k=1$ at the end. Therefore, the higher order
contributions might in principle be greater than the lower order
ones. However, by analyzing the expansion of $Z_0$ we observe that higher
order terms contain $a_k$ coefficients with larger $k$. Moreover, at
least for polynomial potentials, a scaling argument shows that 
the coupling constant dependence of $a_k$ has a hierarchy fixed by $k$
i.e. for larger $k$ one gets more powers of the inverse coupling
constant (see e.g. the expressions given in \eq{f4a0} and
\eq{f4a2}). Therefore, the expansion in $\k$ can be reorganized as an
expansion in the inverse coupling constant. 

\section{Triviality \label{s4}}

In section \ref{s2}, we found that to have a well defined non-Gaussian
path integral in the kinetic expansion the bare coupling constant $\l$
should obey \eq{flow1}. Therefore, $\l$ vanishes for $p>2$ in the
continuum limit as $\L\to\infty$, and one may wonder whether the
theory becomes free in this limit. Although the interactions vanish
classically, this is not an obvious question in quantum theory.
Indeed, there is no sign of triviality in the kinetic expansion
studied in section \ref{s2}.  

Let us try to analyze the situation in usual perturbation theory with
the interaction potential \eq{vv} (so $p$ is assumed to be an
integer). Without loss of of generality we take $p>2$ and focus on the
connected graphs. It is well known that
infinities generically arise in evaluating Feynman diagrams, however
in our case these can be suppressed by the bare coupling constant
since it vanishes like \eq{flow1} as one removes the cutoff $\L$. 
Consider a connected graph with $I$ internal lines, $E$ external
lines, $N$ vertices and $L$ loops. We have the following topological
identities: 
\be\label{top}
2I+E=Np, \hs{10} L=I-N+1.
\ee
The superficial degree of divergence $D$ of this diagram is given by
(for each loop the momentum space integration measure contributes $n$
units, and each internal line propagator adds $-2$)
\be
D=nL-2I.
\ee
Using \eq{top} to eliminate $L$ and $I$ one gets
\be
D=-\fr{nE}{2}-2I+n+\fr{Nn(p-2)}{2}.
\ee
The order of the diagram is $N$ and the degree of suppression $S$ coming
from the cutoff dependence of $\l$ in \eq{flow1} can be found as
\be
S=\fr{Nn(p-2)}{2}.
\ee
Comparing the two numbers we find that $S>D$ . Therefore, as $\L\to\infty$,
$\l^{N}$ term becomes more singular than the outcome of the momentum space
integral and the final result for the diagram should tend to
$0$. According to this naive power counting, the only non-vanishing
graph is the tree level two point function, i.e. the quantum
theory is actually free. It is important to emphasize that 
this is not a formal proof of triviality since
the calculated degree of divergence $D$ arises only from regions of
momentum space in which all internal $n$-momenta go to infinity
together. It is known that additional divergences can also show up when
the momenta belonging to some subgraph tend to infinity. Therefore,
$D$ does not necessarily give the actual degree of divergence of a
diagram. However, the above power counting is suggestive and indicates
triviality.  

Let us now try to use the alternative expression \eq{zk} in evaluating
the generating functional. From \eq{z0e} one finds 
\be\label{zint}
Z_0[f]=\exp\,\left[\int\,\fr{a_2}{a_0}\,f(x)^2\,\, d^n x \right]
\times\,Z_{int}[f],
\ee
where $Z_{int}[f]$ has an expansion of the form
\be\label{zint2}
Z_{int}[f]=\exp\,\left[\sum_{k=2}^{\infty} \e^{k-1}A_k
  \int\,f(x)^{2k}\,\, d^n x \right]
\ee
and $A_k$ are finite coefficients which can be determined in terms of
$a_k$, e.g. $A_2=a_4/a_0-a_2^2/(2a_0^2)$. Using \eq{zint} and
\eq{id1} in \eq{zk}, we get      
\bea
Z[J]&=&\exp \left[-\fr{1}{2}\int J\Box^{-1}J\right]\times \nn\\
&&Z_{int}\left. \left[\fr{\d}{\d K}\right]\, e^{\left[\fr{1}{2}
    \int K\,\D\,K \right]}\,\right|_{\Box K= J} \label{zz}
\eea
where 
\be\label{pd}
\D(x,y)=\Box P_2 (x,y).
\ee
From \eq{zint2} the second line of \eq{zz} is equivalent to the
generating functional of a scalar field $\varphi$  having the
interaction potential $V_{int}$, 
\be\label{vp}
V_{int}=\sum_{k=2}^{\infty} \e^{k-1}A_k \, \varphi^{2k},
\ee
and the propagator $\D$ (here $K$ plays the role of 
the external current coupled to this hypothetical scalar $\varphi$). 
Therefore, it can be evaluated order by
order using the well known perturbation theory techniques, i.e. by
calculating the Feynman diagrams corresponding the interaction
potential \eq{vp}. The only difference is that an extra $\Box^{-1}$
factor should be attached to each {\it external} line since one should set
$\Box K= J$ at the end. 

Consider a connected graph in $\varphi$ theory with $I$ internal
lines, $E$ external lines, $N_k$ vertices of type $\varphi^{2k}$ and
$L$ loops. For this diagram we have
\be\label{top2}
2I+E=\sum_{k}N_k(2k),\hs{8} L=I-N+1,
\ee
where $N=\sum_{k}N_k$ is the total number of interaction
vertices. Upto a finite constant prefactor, the corresponding
truncated function becomes 
\be\label{amp}
G\,\sim\,S\,\prod_{m=1}^{N}\,\prod_{l=1}^{I}\,\int\,d^n
p_l\,\d(m)\,\D(p_l),
\ee
where $\d(m)$ represents the Dirac delta function at the vertex $m$,
$\D(p)$ is the momentum space propagator which can be calculated
from \eq{bp} and \eq{pd} as
\be\label{mp}
\D(p)=\fr{1}{p^{-2}+\fr{2a_2}{a_0}},
\ee
and $S$ is the contribution of the cutoff dependent coupling constants
in \eq{vp}:
\be\label{s}
S=\prod_{k}\e^{(k-1)N_k}=\L^{-n\sum_k(k-1)N_k}.
\ee
Note that the asymptotic behavior of $\D$ is very different than the
propagator of a conventional scalar field and we have 
\be\label{ineq2}
\D(p)<\fr{a_0}{2a_2}.
\ee
Using this inequality in \eq{amp} (and ignoring the finite parts) one
finds  
\be\label{ara}
G\,<\,S\,\prod_{m=1}^{N}\,\prod_{l=1}^{I}\,\int\,d^n
p_l\,\,\d(m). 
\ee
The delta functions impose linear relations among the internal momenta
which reduce the number of independent momentum variables to the
number of loops (of course one delta function survives which
ensures conservation of external momentum). Moreover each loop integral in
\eq{ara} diverges like $\L^{n}$ and thus  
\be
G< S\,\L^{nL},
\ee
which gives an upper bound for the {\it actual} degree of divergence of
this diagram. Solving for $I$ and $L$ from \eq{top2}, and using \eq{s}
one finally obtains 
\be\label{58}
G<\L^{n(1-E/2)}.
\ee
Eq. \eq{58} shows that all graphs which have more than two  external lines
vanish and all corrections to the propagator are bounded in the
continuum limit. Therefore, the theory becomes {\it trivial} as one
removes the cutoff.  

\begin{figure}[htb]
\begin{center}
\centerline{\includegraphics[width=6.0cm]{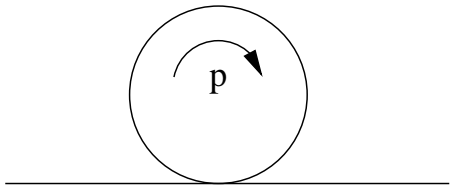}}
\end{center}
\caption{The first order correction to the propagator which arises
  from $\e\,A_2\,\varphi^4$ term in the potential \eq{vp}. The loop integral,
  which is calculated with the propagator $\D$ in \eq{mp},  
  diverges like $\L^n$. This divergence is compensated by the $\e$
  factor and the result becomes finite in the 
  $\L\to\infty$ limit. \label{fig1}}   
\end{figure}

A few comments are now in order. Although corrections to the
propagator are all finite, it is not clear if the sum of these terms 
would converge or if the final result would correspond to a physical
particle. The first term in this series was actually calculated
in \eq{oneloop}. This corresponds to a one-loop diagram in $\varphi$
theory which arises from the first term in the  potential 
in \eq{vp} (see figure \ref{fig1}). The tree level contribution of the
same term to the four-point function is suppressed by the cutoff
dependence of the  ``coupling constant'' in $V_{int}$. Note that in
the second line of \eq{oneloop} the propagator $P_2$ is attached to
the external lines instead of $\D$. This is because, as pointed out
above, the external lines should be multiplied by $\Box^{-1}$ in
obtaining $Z[J]$ and we have $\Box^{-1}\D=P_2$. 

The above argument of triviality also works for non-polynomial
potentials which cannot be analyzed in usual perturbation theory. The
only requirement is that the cutoff dependences of the bare parameters
should be fixed as in \eq{flow1} to have a well defined non-Gaussian
path integral without the kinetic term. In general \eq{flow1} is not
valid for the bare parameters obtained in usual perturbation theory;
for instance in $\l\f^4$ theory in 4-dimensions there is a one loop
logarithmic $\L$ dependence in $\l$. So, the triviality argument
{\it is not} directly applicable to perturbatively treated
self-interactions.   

Returning back the kinetic expansion discussed in section \ref{s2},
one may conclude that the infinities arise simply because locality is
ignored in calculating the zeroth order contribution $Z_0[J]$ and,
oppositely, the corrections demand locality (this viewpoint was
emphasized in \cite{st6}). Therefore, to be able to use \eq{fullz}
one my try to find a way of resolving this dilemma. One possibility
is to set up an RG flow starting from IR as discussed in footnote
\ref{ftrg}. It would be interesting to verify triviality directly from
\eq{fullz} by using this RG flow, so that the locality can be
recovered step by step starting from IR.  

\section{Conclusions}

In this paper, an alternative expansion scheme is obtained
in self-interacting scalar field theories which
uses the elements of kinetic expansion and usual perturbation theory
together. The central object in this framework is the non-Gaussian
functional integral of the potential. These integrals naturally arise
in the kinetic expansion, which, however, is plagued by the presence of
infinities. Kinetic expansion is ill defined since locality is
completely neglected in the zeroth order configuration around which
there is a local expansion. To resolve the problem we introduced a
Lagrange multiplier to separate the kinetic and the potential terms in
the path integral. In this way, we are able to invert the kinetic term
while the potential can be integrated out non-perturbatively but order by
order in a cutoff parameter yielding a series expansion for the
generating functional. Although we just consider a real scalar field, 
it is not difficult to generalize the formulas for coupled
multiple scalars. For instance, the analog of the one dimensional
integral \eq{z=} is a multi-dimensional one fixed by the number of
scalars in the theory.     

Using this new scheme, we show that if the bare parameters in the
potential are chosen to have a well defined non-Gaussian path integral
then all graphs except the two-point function vanish in the continuum
limit. Therefore, the theory becomes trivial when the cutoff is
removed. Note that the bare couplings are fixed ``kinematically'',
i.e. to define a proper integration measure.  For polynomial
potentials it should be possible to give a proof of triviality in
usual perturbation theory as suggested by the naive power
counting discussed in section \ref{s4}. However, it is difficult to
see triviality in the kinetic expansion.  

Changing the potential only modifies the constants $a_m$ (i.e. their
dependences on the parameters of the theory) and the other steps in the
calculations remain unaffected.  This indicates that  the
self-interaction  potential simply plays the role of a weight function
in the path integral. Note that in Euclidean space the convergence of
the generating functional  (\ref{z1})  should be more rapid for 
``larger'' interaction potentials  since  contribution of  each  field
configuration in  the functional  integral   would  decrease   due  to
the $\exp(-\int  V)$ factor. This can be observed in our approach since 
the path integral containing the potential reduces to product of 
ordinary integrals. However, this is contrary to what we see in usual 
perturbation theory; for example in four-dimensions although $\l\f^4$
theory is perturbatively tractable $\l\f^6$ is not.  

It would be interesting to extend the present work in different
directions. Firstly, it is possible to combine the above formalism
with the usual perturbation theory. This can be achieved by taking
{\it all} bare quantities  in the Lagrangian (the wave-function, the
coupling constants etc.) as provided by the perturbative
renormalization theory and use them in the expression \eq{zk} or
\eq{zson}. This may be helpful in addressing puzzles like the
hierarchy problem. Secondly, one may consider non-Gaussian path
integrals for self-interacting fermions. Finally, it would be
interesting to study scalars coupled to gauge fields and, especially, to
analyze the mechanism of spontaneous symmetry breaking using
non-Gaussian path integrals.     

\begin{acknowledgments}
I would like to thank N.S. Deger, T. Rador and T. Turgut
for useful discussions. I also would like to thank M. Hortacsu and
K. Scharnhorst for bringing out the earlier work on the subject to my
attention.  
\end{acknowledgments}

\end{document}